\documentclass[12pt,a4paper]{article}
\usepackage{amsmath}
\usepackage[dvips]{graphicx}
\input epsf
\usepackage{times}

\begin{document}
\begin{titlepage}
\title{The effects of reflective scattering  in the spin correlation parameter  in top quark production at the LHC}
\author{S.M. Troshin, N.E. Tyurin\\[1ex]
\small  \it Institute for High Energy Physics,\\
\small  \it Protvino, Moscow Region, 142281, Russia} \normalsize
\date{}
\maketitle
\begin{abstract}
The specific effects of the reflective scattering (antishadowing) related to the spin correlations in
the top quark production at the LHC are considered. Account for those effects is important in the  searches
for the signatures of the extra dimensiones.  It is shown
that significant spin correlations could arise  at the LHC
energies due to reflective scattering and those can affect the signals of the extra dimensions.

\end{abstract}
\end{titlepage}
\section*{Inroduction}
The studies of multiparticle production can reveal high degrees of correlations among the spins of  produced
particles, i.e. particles would be produced with  aligned spins. 
Chou and Yang   came to  conclusion on existence of such correlations of particle spins on the grounds
of their geometrical \cite{chyn}.  Similar conclusion on  spin correlations   at very high
energies has been made later on the base of the reflective scattering mechanism \cite{meplb}. This mechanism
will be discussed later in the context of the top quark priduction at the LHC.

At the LHC colliding proton beams are not polarized and the only way to perform spin-dependent mesurements 
 is to determine the spin directions of final particles  via their decay products kinematics. 
One such possibility to detect global spin correlations
\cite{meplb} is related to  $\Lambda$-hyperons production and their spins measurements.
 Another possibility, almost exlusively appropriate to the LHC, is
related to the top quark production. This latter process has high statistics and also allows spin measurements through the restoration of the 
decay products kinematics. 

At the LHC, the Standard Model (SM) dominating mechanism of the top quark pairs
 production is due to gluon fusion. Spin correlations in the top-antitop production in the SM have been calculated in
\cite{mahlon}.   The main interest in their studies is related to sensitivity of those spin correlations
 to the presence of the
extra dimensions \cite{arai}. 

In present note we briefly report the known results on these correlations and discuss the possible
effects due to the reflective scattering mechanism. 
We estimate the size and sign of these effects for the LHC.

\section{Spin correlations at the LHC in the Standard Model}
As it was noted, in perturbative QCD the top-antitop pairs are produced at the LHC mainly in the gluon
 fusion processes \cite{mahlon}, $gg\to t\bar t$, and top decays
prior to the hadronization.
This feature allows one to determine top quark spin direction by measuring corresponding angles of the decay products.
The decay modes in the SM consist of leptonic and hadronic ones:
\begin{equation}\label{dec}
 t\to bW^+\to bl^+(l=e,\mu,\tau)\nu_l,\, bu\bar d,\, bc\bar s.
\end{equation}

The direction  of the top quark spin can be reconstructed via its decay products kinematics.
The differential decay rates can be written in the top quark rest frame in the  form:
\begin{equation}\label{direc}
 \frac{1}{\Gamma}\frac{d\Gamma}{d\cos \theta_f}=\frac{1}{2}(1+\kappa_f \cos \theta_f),
\end{equation}
where $\Gamma$ is the partial decay width of the particular channel, $\kappa_f$ is the top-spin analysing power and
$\theta_f$ is the angle between the top-spin axis and direction of momemtum of the particle $f$. The values
of the analysing power for different particles have been calculated in the Standard Model (cf. e.g.  \cite{arai}).
The spin correlation parameter defined as
\begin{equation}\label{corr}
C_{t\bar t}=\frac{\sigma(t_\uparrow\bar t_\uparrow)+\sigma(t_\downarrow\bar t_\downarrow)-\sigma(t_\uparrow\bar t_\downarrow)-
\sigma(t_\downarrow\bar t_\uparrow)}{\sigma(t_\uparrow\bar t_\uparrow)+\sigma(t_\downarrow\bar t_\downarrow)+\sigma(t_\uparrow\bar t_\downarrow)+
\sigma(t_\downarrow\bar t_\uparrow)}
 \end{equation}
can be extracted from the double differential angular distribution of the respective decay products with angles
$\theta_1$ and $\theta_2$:
\begin{equation}\label{dble}
\frac{1}{\sigma}\frac{d^2\sigma}{d\cos \theta_1 d\cos \theta_2}=[1-C_{t\bar t}\cos \theta_1 \cos \theta_2]/4.
 \end{equation}
In Eq. (\ref{corr})  the $\sigma(t_\alpha \bar t_{\tilde \alpha})$ are the cross-sections of the top-antitop pair production with
respective spin directions. In the SM spin correlation parameter has been calculated for the LHC at the lowest order in $\alpha_s$ and the values
$C_{t\bar t}=0.30-0.31$ have been obtained.  The details can be found in \cite{mahlon}, it should be noted that numerical values
of spin correlations  depend on the particular parametrizations of the parton distribution functions.  Besides, the value of the top-quark
spin correlations can be affected by several mechanisms considered in the next sections.
\section{Extra dimensiones and the spin correlations}
The possibility of the presence  extra spacial dimensions in addition to the usual 3+1 dimensional
manifold has been considered in \cite{add,rs}.  The main purpose for the introduction of the new
dimensions was solution of the hierarchy problem, i.e. huge difference in scales of electroweak and
gravitational interactions. The 4-dimensional Planck scale $M_{Pl}$ is connected
with the fundamental scale in $4+n$ dimensiones $M$ and radius of the compactified $n$ extra dimensiones $R$
by the following relation  \cite{add}: 
\begin{equation}
 M_{Pl}=M(MR)^{n/2}.
\end{equation}
When the radius $R$ is large ($R\sim 0.1$ mm for $n=2$) the value of $M$ can be around 1 TeV and it allows one to solve the
gauge hierarchy problem.
In another theory called 5-dimensional warped geometry theory \cite{rs}  our space
 is a five-dimensional anti de Sitter space and the elementary particles
 except for the graviton are localized on a (3 + 1)-dimensional brane or branes.
We will not discuss the difference in these approaches, it is important to note
that in the effective theory graviton propogating in the bulk can be expressed
as an infinite tower of Kaluza-Klein (KK) spin-2 gravitons.
Thus, in the 4-dimensional effective theory KK-gravitons interact with the SM fields.
This interactions change the total cross-section of the top quark production (increase it by factor 2 for
the values of $M$ around 0.5-1 $TeV$). It also leads to significant changes in the spin correlation parameter
decreasing it from 0.3 in the SM to zero or even negative value in this region of $M$. This is due to the spin-2 nature
of the KK gravitons. For the other values of $M$ the
role of interactions with KK-gravitons is minimal \cite{arai}. 
There is another global dynamical mechanism which should be taken into consideration. 
It 
is related to the reflective scattering.
\section{Effects of reflective scattering in the spin correlations}
The idea that reflective scattering can lead to the global spin correlations of the final particles has been
considered in \cite{meplb}. In this section we apply qualitative conclusions of the above paper to the top-quark production
and estimate numerical value of the respective spin correlation parameter.
The notion of reflective scattering is related to appearance of the phase factor $e^{i\pi}$ in the elastic scattering $2\to 2$ matrix element 
$S(s,b)$. We discuss here for simplicity the case of pure imaginary scattering amplitude.  The negative sign of $S(s,b)$ is a manifestation
of the fact that the elastic scattering amplitude $f(s,b)$ is greater than $1/2$, note that $S(s,b)=1-2f(s,b)$.
Reflective scattering is a
result of unitarity saturation, when partial amplitude tends to unity at $s\to\infty$.
Such saturation  realised in  the $U$--matrix or rational form  of the amplitude unitarization
at high energies. Thus, as it was noted already, the saturation of unitarity is characterized by the fact that beyond 
some threshold  energy value the elastic scattering matrix element $S(s,b)$ at $b=0$ becomes negative (i.e. $S(s,b=0)\to -1$ at $s\to \infty$) and
the inelastic overlap function
\[
\eta(s,b)\equiv\frac{1}{4\pi}\frac{d\sigma_{inel}}{db^2}
\]
starts to develop a  peripheral impact parameter dependence since 
\[
\eta(s,b)=f(s,b)(1-f(s,b)),
\]
since $f>1/2$.  At the LHC energies
this peripheral profile with a maximum at $b=R(s)$, where $R(s)$ is
the  radius of reflective scattering determined by the relation $S(s,b=R(s))=0$, should  become quite noticeable.
The usual exponential (eikonal) form of unitarization does not lead to
such a dependence on the impact parameter. The difference in the impact parameter dependencies
is illustrated in Fig. 1.

The region around the values of the impact parameter  $b=R(s)$ has the
highest probability  of the multiparticle production given by the relation
 $P_{inel}(s,b)=4\eta (s,b)$.
  Mechanism of reflective scattering leads to suppression  of particle
production at small impact parameters and the main contribution to
the integral  multiplicity $\langle n\rangle (s)$ comes from the region of
$b\sim R(s)$.
Thus, due to peripheral impact parameter dependence of the inelastic overlap function the
secondary particles will be mainly
 produced at the impact parameters $b\sim R(s)$ and this will lead to imbalance
 between orbital angular momentum in the initial and final states since the most secondary produced particles
  would carry  large orbital
 angular momentum.
To compensate this imbalance in the orbital momentum 
  the spins of the secondary particles should   be 
  correlated, i.e. aligned.  Such correlations are expected to appear at the energies where the reflective scattering occures.
\begin{center}
\begin{figure}[hbt]
\hspace*{2cm}\epsfxsize= 100  mm  \epsfbox{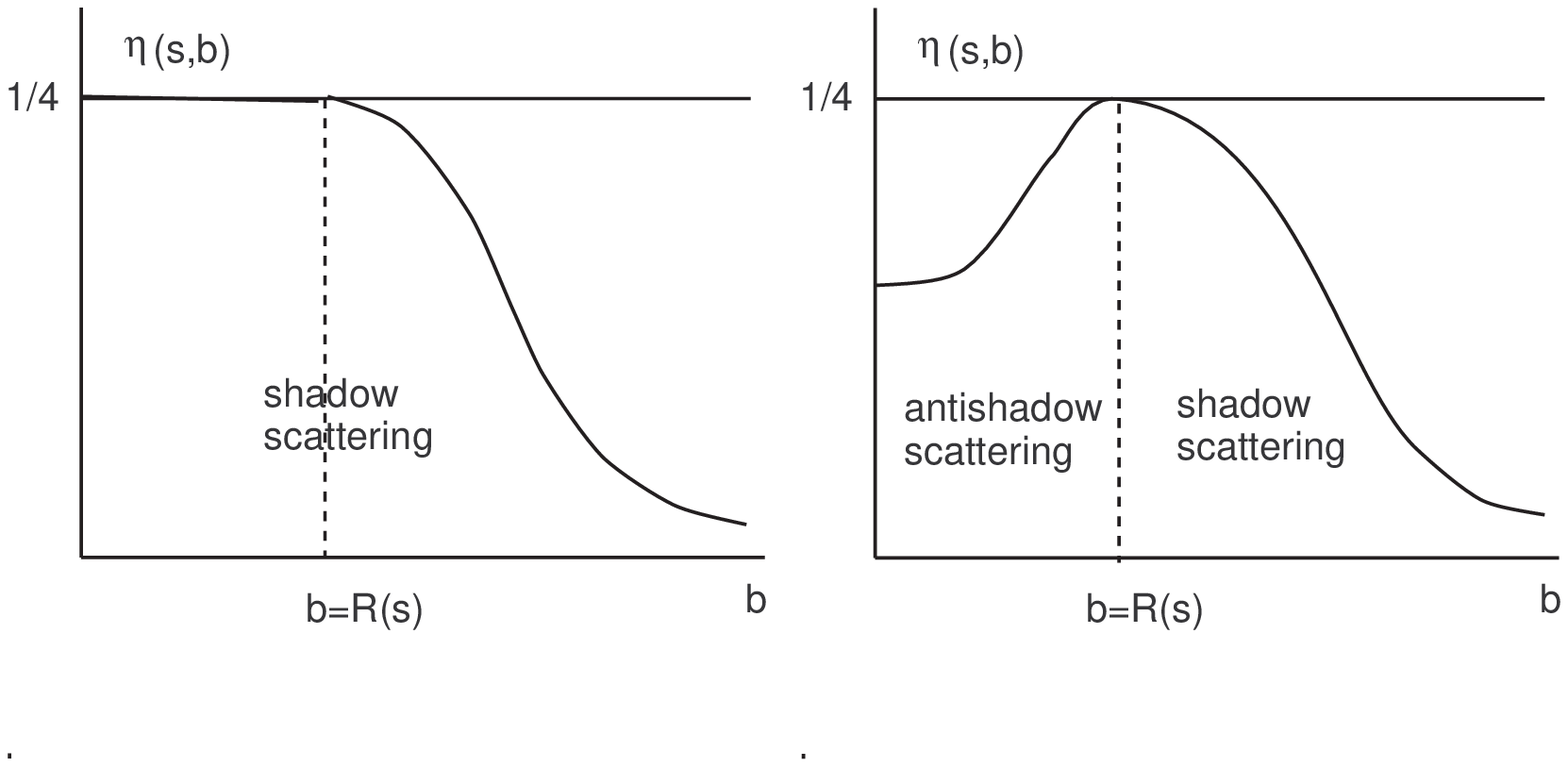}
\caption{\small{Impact parameter dependence of the inelastic overlap
function
 in the standard unitarization scheme (left panel) and in the unitarization scheme
 with reflective scattering presence (right panel).}}
 \end{figure}
\end{center}
Such considerations are appropriate for all inelastically produced  particles, i.e. the spin correlations have a global nature. Despite that,
it is difficult to determine spin directions  of the most final particles. In this sense, the top quark is
one of exclusions (another one is $\Lambda$-hyperon productions for example),  and as it was already noted, 
its spin can be directly measured through its decay.
On the other side, production of top-quark pairs  is an inelastic process and it should be affected by unitarity constraints as well. 
The imbalance in the orbital momentum will lead to spin correlations in the top-antitop
production at the LHC where the reflective scattering should take place.

Thus on the grounds of unitarity saturation, the value of spin correlation parameter $C^r_{t\bar t}$ should be proportinal to
the relative imbalance of the orbital
angular momentum and to estimate it  we make the assumption and just equate these two quantities, i.e.
\begin{equation}\label{correlp}
 C^{r}_{t\bar t}=-\frac{\Delta L(s)}{L(s)},
\end{equation}
where $L(s)$ is the orbital angular momentum of the initial state in the region of reflective scattering and
$C^{r}_{t\bar t}$ is the spin correlation parameter arising as a result of the reflective scattering.  As it was discussed earlier,
 the spin correlation parameter
has the same value
  for all other particles, but spin of $t$-quark can be determined due to its decay. 
The negative sign  in Eq. (\ref{correlp}) appears due to the fact that imbalance in orbital angular momentum is compensated
by the spin correlations of the final particles in order to provide conservation of the total angular momentum. The numerical
value of this spin correlation parameter can be calculated using inelastic overlap function (cf. Fig. 1)
\begin{equation}\label{correlpn}
 C^{r}_{t\bar t}=\frac{\int_0^{R(s)}[1-P_{inel}(s,b)]bdb}{\int_0^{R(s)}bdb}.
\end{equation}
Since $P_{inel}(s,b)=4\eta(s,b)$ and $S^2(s,b)=1-4\eta(s,b)$ we will have the following relation valid in the case
when reflective scattering is included (i.e. when saturation of unitarity is assumed at asymptotic energies):
\begin{equation}\label{correlpn1}
 C^{r}_{t\bar t}=\frac{2}{R^2(s)}\int_0^{R(s)}S^2(s,b)bdb,
\end{equation}
To obtain the numerical results specific models and parameter values should be used. We apply the model developed
in \cite{phrev}. It is based on the rational form unitarization where $S(s,b)$ has the form \cite{intj}
\begin{equation}\label{ru}
S(s,b)=\frac{1-U(s,b)}{1+U(s,b)}.
 \end{equation}
  The explicit form of $U(s,b)$
and values of the model parameters are described e.g.  in \cite{dpe}. The numerical values for the
spin correlation parameter can be calculated according to Eq. (\ref{correlpn1}): 
\[
  C^{r}_{t\bar t}\simeq 0.2\,
( \sqrt{s}=7\, TeV)\,\, \mbox{and}\,\,
C^{r}_{t\bar t}\simeq 0.4\,
( \sqrt{s}=14\, TeV).
\]

Thus, the  reflective scattering mechanism at the LHC energies can significantly affect the expected values of the spin correlation
parameters, e.g.  it can compensate small values of spin correlation parameter predicted due to possible presence of the extra
dimensions providing not accounted significant background. There is a way to separate contribution to the spin correlation parameter
coming from SM model gluon-gluon fusion, contribution of the virtual KK-gravitons and reflective scattering. Reflective scattering
has a global nature and will lead e.g. to the value $C_{tt}\simeq 0.2$ at 7 TeV while former processes would provide zero value for
this parameter due to independence of the double-parton scattering. Inderectly, this can be tested by measuring $C_{\Lambda\Lambda}$ for example.
It should be noted that  spin correlation parameter in the top-antitop
production in the helicity basis has already been
 measured at the LHC by ATLAS  \cite{atlas} (with the value $0.40^{+0.09}_{-0.08}$). 
This number allows one to make a
conclusion on the agreement with the SM predictions. However,  one also can speculate 
on the possibility of the presence of the another mechanism, namely the
presence of  contribution of the KK-gravitons  which is corrected by the reflective scattering effects. 

\section*{Conclusion}
We discuss here the spin correlation parameter in the top quark
production at the LHC. This observable is sensitive to  the effects of the extra dimensions and
was considered in the literature as a clear  signal of the KK-gravitons existence on the brane.
At the same time there is an expectation  of additional contribution into the spin
correlations in top quark production at the LHC energies resulting from the possible existence of the prominent
reflective scattering mechanism. This can provide significant background
for the detection of  the new physics effects associated with the  extra dimensions 
 and should be taken into account under  the correct interpretations of the experimental results. 
The effects of reflective scattering can also  be verified by the spin correlation parameter 
measurements in the $\Lambda$-hyperon production as it was discussed in \cite{meplb}.
\section*{Acknowledgements}
We would like to thank R.~Chierici, V.~Kachanov, D.~Konstantinov, V.~Petrov, R.~Tenchini and A.~Uzunian for the 
 discussions of the results  and comments.

\end{document}